\documentclass[aps,prb,twocolumn,groupedaddress,nofootinbib]{revtex4-2}
\usepackage{amsmath,amssymb,graphicx,bm}
\usepackage{siunitx}
\usepackage{physics}
\usepackage{xcolor}
\usepackage{graphicx}

\begin{document}

\title{Interaction-Induced Spiral Motion and Breathing Dynamics of N\'eel Skyrmions in Ferromagnetic Thin Films}

\author{M. Al Busaidi, F. Al Ma'Mari and R. Sbiaa*}
\affiliation{Department of Physics, College of Science, Sultan Qaboos University, P.O. Box 36, PC 123, Muscat, Oman}
\date{\today}

\email{rachid@squ.edu.om}  

\begin{abstract}
Magnetic skyrmions exhibit particle-like stability and rich dynamical behaviour arising from their topological nature, making them promising building blocks for future spintronic devices. In this work, we investigate the interaction dynamics of two N\'eel-type skyrmions in an ultrathin ferromagnetic film through a combined analytical and numerical approach. 
Full micromagnetic simulations reveal that when the initial separation exceeds twice the skyrmion radius, the pair undergoes a repulsive interaction leading to an outward spiral trajectory whose radial and angular components depend sensitively on the Gilbert damping, consistent with particle-like models of skyrmion motion. 
The simulations further show a two-component oscillatory behavior in the out-of-plane magnetization: a fast intrinsic breathing mode superimposed on a slower interaction-induced modulation. Using Thiele collective-coordinate model, an expression for the radial drift, angular decay, and logarithmic growth of the separation distance is derived. The analytical predictions show excellent agreement with numerical results, confirming the exponential form of the long-range interaction potential. A continuum micromagnetic analysis of the skyrmion tail explains the origin of the exponential decay and its role in governing the interaction strength. Together, these findings provide a unified framework for understanding spiral motion, breathing dynamics, and long-range repulsion in interacting skyrmion systems, offering 
insights relevant for multi-skyrmion information carriers and collective skyrmion-based devices.

\end{abstract}

\maketitle

\section{Introduction}

Magnetic skyrmions are nanoscale, topologically protected spin textures that arise in chiral and ultrathin ferromagnets due to the competition between exchange interaction, anisotropy, and Dzyaloshinskii--Moriya interaction (DMI)~\cite{Fert2017_NatRevMat_Skyrmions,Nagaosa2013_NatNanotech}. Their particle-like stability, small size, and efficient current-driven mobility make them promising candidates for next-generation spintronic devices, including racetrack memories~\cite{Tomasello2014_SciRep_SkyrmionRacetrack,Song2017_SkyrmionRacetrack,Bessarab2018_SkyrmionLifetime,Huang2020_NatElectron_SkyrmionSynapse,He2023_NanoLett_SkyrmionMemory,Pham2024_Science_SkyrmionSAF}, neuromorphic architectures~\cite{Zhang2018_NanoLett_SkyrmionNeuron,Song2020_NatElectron_SkyrmionSynapses,Yu2020NanoscaleAdv,Lone2024_Nanoscale_MultilayerSpintronic,Verma2025_APL_FieldFreeSkyrmionNN,Aslam2025_ACSAELM_SkyrmionMemristor}, and logic components~\cite{Zhang2015_SciRep_SkyrmionLogic,Mankalale2019_IEEE_SkyLogic,Yu2020NanoscaleAdv,Sisodia2022_PRApp_SkyrmionLogic,MousaviCheghabouri2024_AdvTheorySimul_SkyrmionLogic,Belrhazi2024_ACSApplMater_SkyrmionLogic,Pan2024_PRResearch_SkyrmionQubits}. Because skyrmions behave as emergent quasiparticles, understanding their dynamical interaction laws is essential for predicting collective phenomena, collision processes, and the formation of composite states such as biskyrmions!~\cite{Yu2014_NatCommun_Biskyrmions,Song2023_SciAdv_Biskyrmions,AlSaidi2025_MatTodayE_SkyrmionBiskyrmion} or skyrmion bags~\cite{Tang2021_NatNanotech_SkyrmionBundles,Bo2023_PRB_SkyrmionBags,Wang2023_PRB_ParticleContinuumSkyrmions}.

The interaction between skyrmions plays a central role in their collective dynamics. While isolated skyrmions exhibit gyrotropic motion dominated by the Magnus force, skyrmion pairs experience additional forces arising from their mutual distortion and long-range tail overlap~\cite{Capic2020_JPhysCondMat_SkyrmionInteraction,Ross2021_SkyrmionInteractions}. Several studies have analyzed the effective skyrmion--skyrmion potential, revealing regimes of repulsion, attraction, or bound-state formation depending on chirality, polarity, and DMI strength~\cite{Capic2020_JPhysCondMat_SkyrmionInteraction,Ross2021_SkyrmionInteractions,Gobel2019_Biskyrmions}. These interactions are also closely tied to the internal eigenmodes of skyrmions---most notably the breathing mode, in which the skyrmion radius exhibits high-frequency oscillations~\cite{Kim2014_PRB_BreathingSkyrmionDots,McKeever2019_PRB_BreathingDynamics,Garanin2020_BreathingMode,Mochizuki2012_PRL_SkyrmionSpinWave}. Such breathing dynamics not only influence the stability of individual skyrmions but also affect the radial motion and collision outcomes of skyrmion pairs~\cite{Kim2014_PRB_BreathingSkyrmionDots,McKeever2019_PRB_BreathingDynamics}.

Despite extensive theoretical efforts, a complete and unified description of the skyrmion--skyrmion interaction dynamics---linking analytical models to full micromagnetic behavior---remains limited. The Thiele equation provides a powerful reduced description of skyrmion motion as a rigid particle subject to gyrotropic, dissipative, and interaction forces~\cite{Thiele1973_PRL_DomainMotion,Lin2013_PRB_ParticleModel}. However, it cannot capture internal deformations or high-frequency breathing oscillations, which become significant during close-range interactions~\cite{McKeever2019_PRB_BreathingDynamics,Garanin2020_BreathingMode}. Conversely, continuum micromagnetic theory offers analytical insight into the asymptotic form of the skyrmion tail and the exponential decay of the interaction potential~\cite{Ross2021_SkyrmionInteractions,Capic2020_JPhysCondMat_SkyrmionInteraction}, while large-scale micromagnetic simulations reproduce the full nonlinear dynamics, including spiral trajectories, damping-induced drift, and complex collision behavior~\cite{Shi2023_JPhysD_SkyrmionCollision,Reichhardt2020_arXiv_MagnusClusters,Lin2013_PRB_ParticleModel,Buttner2015_NatPhys_SkyrmionDynamics}.

In this work, we combine three complementary methodological frameworks to study the interaction, dynamics, and stability of two skyrmions in an ultrathin ferromagnetic film:  
(i) full micromagnetic simulations using  Landau–Lifshitz–Gilbert (LLG) equation~\cite{Vansteenkiste2014_AIPAdv_MuMax3},  
(ii) the Thiele collective-coordinate model to describe radial and angular motion analytically~\cite{Thiele1973_PRL_DomainMotion,Lin2013_PRB_ParticleModel}, and  
(iii) a continuum micromagnetic field-theory approach to obtain closed-form expressions for the long-distance skyrmion tail~\cite{Capic2020_JPhysCondMat_SkyrmionInteraction,Ross2021_SkyrmionInteractions}.  
By integrating these approaches, we establish a comprehensive multiscale description connecting microscopic spin interactions to emergent quasiparticle behavior.  

Our results demonstrate that skyrmion pairs with identical polarity and opposite chirality and helicity interact repulsively at distances greater than twice the skyrmion radius, generating spiral trajectories whose radial expansion is strongly dependent on the Gilbert damping parameter, in agreement with previous work on Magnus-dominated particle motion~\cite{Lin2013_PRB_ParticleModel,Ross2021_SkyrmionInteractions}. Moreover, we show that the out-of-plane magnetization exhibits a two-component oscillatory pattern: a fast breathing oscillation superimposed on a slow spiral-induced modulation~\cite{Kim2014_PRB_BreathingSkyrmionDots,Garanin2020_BreathingMode,McKeever2019_PRB_BreathingDynamics,Mochizuki2012_PRL_SkyrmionSpinWave}. Based on Thiele model, we developed an analytical prediction which shows an excellent agreement with micromagnetic simulations, confirming the logarithmic increase of separation distance over time and the exponential form of the interaction potential~\cite{Capic2020_JPhysCondMat_SkyrmionInteraction,Ross2021_SkyrmionInteractions}. This unified framework provides valuable insights into skyrmion--skyrmion collisions, stability mechanisms, and the emergence of breathing dynamics in interacting skyrmion systems.

\section{Methodology}

The theoretical description of skyrmion interactions in this work is built on three hierarchically connected models---the full Landau--Lifshitz--Gilbert (LLG) equation, the Thiele collective-coordinate formalism, and the continuum micromagnetic field theory---each capturing different levels of physical detail~\cite{Fert2017_NatRevMat_Skyrmions,Nagaosa2013_NatNanotech,Ross2021_SkyrmionInteractions}. The starting point is the LLG equation,
\begin{equation}
\frac{\partial \mathbf{m}}{\partial t}
= -\gamma\, \mathbf{m}\times\mathbf{H}_{\mathrm{eff}}
+ \alpha\, \mathbf{m}\times\frac{\partial \mathbf{m}}{\partial t}
\end{equation}
which provides a complete, spatially resolved description of the magnetization dynamics $\mathbf{m}(\mathbf{r},t)$ in ultrathin ferromagnetic films. The parameters $\gamma$ and $\alpha$ are respectively the gyromagnetic ratio and damping constant. The effective field derives from the total micromagnetic energy through
\begin{equation}
\mathbf{H}_{\mathrm{eff}}
= -\frac{1}{\mu_{0} M_{s}}\,\frac{\delta E}{\delta \mathbf{m}}
\end{equation}
and
\begin{equation}
E=E_{\mathrm{ex}}+E_{\mathrm{ani}}+E_{\mathrm{DMI}}+E_{\mathrm{Z}}+E_{\mathrm{dem}}
\end{equation}

The competing interactions give rise to the skyrmion profile. The term $E_{\mathrm{ex}} = A (\nabla \mathbf{m})^{2}$ is the exchange energy that gives rise to the skyrmion profile. The term $E_{\mathrm{ani}} = K_{u} (1-m_{z}^{2})$ is the anisotropy energy that stabilizes the out-of-plane alignment of magnetic moments. $E_{\mathrm{DMI}} = D\!\left[m_{z}\,\nabla\!\cdot\!\mathbf{m}_{\perp}
- \mathbf{m}_{\perp}\!\cdot\!\nabla m_{z} \right]$ is the interfacial DMI which imposes a fixed chirality and sets the handedness of the N\'eel wall, while the demagnetizing energy introduces long-range dipolar interactions~\cite{Fert2017_NatRevMat_Skyrmions,Nagaosa2013_NatNanotech}. 
In this study, the saturation magnetization $M_s$, exchange stiffness $A$, anisotropy constant $K_u$ and iDMI were chosen to be 600 kA/m, 7 pJ/m, 0.65 MJ/$m^3$ and 2 mJ/$m^2$, respectively multilayers~\cite{Fert2017_NatRevMat_Skyrmions,Woo2016_NatMat_RoomTempSkyrmions,AlSubhi2019_CoNi,Sbiaa2016_CoNiCoPt_FMR}.
 Solving the LLG equation fully captures skyrmion deformation mechanisms such as breathing, shrinking, shearing, and skyrmion--skyrmion merging, which become crucial at small separations where the cylindrical symmetry of each skyrmion breaks down~\cite{Shi2023_JPhysD_SkyrmionCollision,Brearton2020_arXiv_SkyrmionInteractions,Gobel2019_Biskyrmions}.

To complement the full-field LLG description, we adopt the Thiele collective-coordinate model, which assumes that a skyrmion behaves approximately as a rigid quasiparticle whose dynamics are dominated by the translational mode. Under this assumption, integrating the LLG equation over space yields the Thiele equation,
\begin{equation}
\mathbf{G}\times\mathbf{v} + \alpha D\,\mathbf{v} + \nabla U(r) = 0,
\end{equation}
where $\mathbf{v}=\dot{\mathbf{X}}$ is the skyrmion velocity, $\mathbf{G}=-4\pi Q\,\hat{\mathbf{z}}$ is the gyrovector reflecting the topology ($Q=\pm1$), and $D$ is the dissipative tensor set by the internal magnetization gradients~\cite{Thiele1973_PRL_DomainMotion,Lin2013_PRB_ParticleModel}. Writing the equation in polar form gives
\begin{align}
-G\dot{\phi} + \alpha D\dot{r} &= F(r) \\
G\dot{r} + \alpha D r\dot{\phi} &= 0
\end{align}
with $F(r)=-U'(r)$ is the skyrmion--skyrmion interaction force. These relations explicitly show that the radial force $F(r)$ drives angular motion through the gyrotropic term, producing spiral trajectories~\cite{Lin2013_PRB_ParticleModel,Reichhardt2020_arXiv_MagnusClusters}. The model is accurate when the skyrmion retains its shape (moderate separations or low velocities) but becomes unreliable when strong deformations occur, such as in short-range attraction ($r \lesssim 2R$) or during skyrmionium formation~\cite{Gobel2019_Biskyrmions,Ross2021_SkyrmionInteractions,Capic2020_JPhysCondMat_SkyrmionInteraction}.

Finally, to determine the static structure, size, and energetic stability of isolated and interacting skyrmions, we use the continuum micromagnetic field theory in which the magnetization $\mathbf{s}(\mathbf{r})$ is treated as a continuous unit vector field~\cite{Ross2021_SkyrmionInteractions,Capic2020_JPhysCondMat_SkyrmionInteraction}. Minimizing the energy functional,
\begin{multline}
E=\int d^{2}r\,\Big[
A(\nabla \mathbf{s})^{2}
+ K_{\mathrm{eff}} (1-s_{z}^{2})+\\
 D\big(s_{z}\nabla\!\cdot\!\mathbf{s}_{\perp}
      - \mathbf{s}_{\perp}\!\cdot\!\nabla s_{z}\big)  
+ E_{\mathrm{Z}}
+ E_{\mathrm{dem}}
\Big]
\end{multline}
yields the Euler--Lagrange equations governing the radial skyrmion profile $s_{z}(r)$ and the characteristic domain-wall width $\Delta=\sqrt{A/K_{\mathrm{eff}}}$. The continuum model also provides the stability condition $D < D_{c}=\frac{4}{\pi}\sqrt{A K_{\mathrm{eff}}}$ and predicts how the skyrmion radius responds to changes in material parameters~\cite{Fert2017_NatRevMat_Skyrmions,Ross2021_SkyrmionInteractions}. This establishes the energetic origin of the interaction potential $U(r)$ used in the Thiele model and explains why only skyrmions with radii close to the equilibrium radius remain dynamically stable. Altogether, these three frameworks form a unified multiscale methodology: the continuum model determines the intrinsic skyrmion structure, the Thiele model captures near-rigid collective motion, and the LLG equation resolves the full nonlinear dynamics, including deformation, breathing, repulsion, attraction and skyrmionium formation~\cite{Gobel2019_Biskyrmions,Shi2023_JPhysD_SkyrmionCollision,Brearton2020_arXiv_SkyrmionInteractions}.

\section{Results and Discussion}

\subsection{Numerical Simulation}

In numerical simulations, we investigate the dynamics of two interacting skyrmions that share the same polarity but possess opposite chirality. Both skyrmions have an identical radius of $R=30~\mathrm{nm}$, and the initial separation between their centers is set to $r=64~\mathrm{nm}$. The interaction is examined in the absence of any external stimuli, such as magnetic fields or spin-polarized currents, so the initial velocity for both of them is the same. In the absence of external fields, the dynamical behavior of skyrmions depends strongly on how many of them are present. A single skyrmion is influenced only by its initial configuration and its interaction with the system boundaries, whereas when multiple skyrmions exist, their mutual interactions must also be taken into account. The simplest non-trivial case is a system containing two skyrmions, which is the situation considered in this study. Our initial observation is that, within this regime, the skyrmions experience a repulsive interaction, causing their separation distance to increase over time, in agreement with previous studies of skyrmion--skyrmion repulsion at large separations~\cite{Capic2020_JPhysCondMat_SkyrmionInteraction,Reichhardt2020_arXiv_MagnusClusters,Shi2023_JPhysD_SkyrmionCollision,Ross2021_SkyrmionInteractions}.

\begin{figure}[b!]
    \centering
    \includegraphics[width=1.0\linewidth]{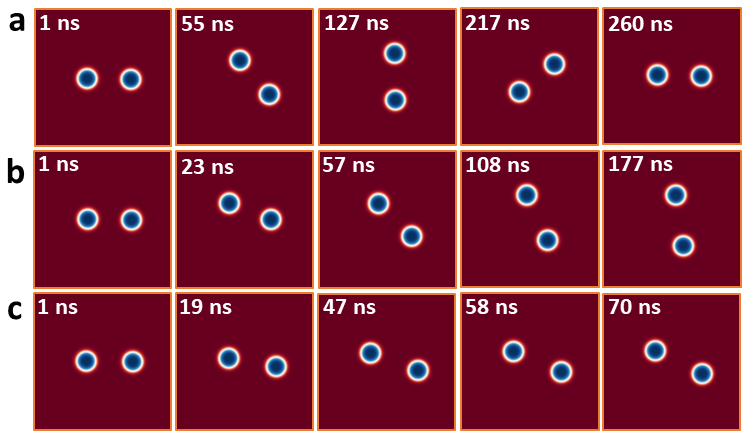}
\caption{
Spiral motion of two skyrmions for different damping values. (a), (b) and (c) are snapshots taken at different times for $\alpha = 0, 0.02$ and $0.1$, respectively}
    \label{fig:FIG1.png}
\end{figure}

Fig.~\ref{fig:FIG1.png} shows the dynamics of two interacting skyrmions initially separated by a distance 
$r > 2R = 64~\text{nm}$. In the presence of damping, the pair follows a spiral trajectory rather than a 
straight or circular path, arising from the interplay between the Magnus force and the repulsive 
interaction associated with their identical polarity, as captured by particle-level descriptions of skyrmion dynamics~\cite{Thiele1973_PRL_DomainMotion,Lin2013_PRB_ParticleModel,Reichhardt2020_arXiv_MagnusClusters}.

In Fig.~\ref{fig:FIG1.png}(a), the undamped limit ($\alpha = 0$) produces a motion that is almost purely circular, dominated by the gyrovector term; in an ideal lossless system, the skyrmions would continue rotating indefinitely. At intermediate damping ($\alpha = 0.02$), shown in Fig.~\ref{fig:FIG1.png}(b), the rotation persists for a longer duration and the increase in separation occurs more gradually. In contrast, Fig.~\ref{fig:FIG1.png}(c) shows that strong damping ($\alpha = 0.1$) rapidly suppresses the angular motion while significantly enhancing the outward radial expansion of the pair, consistent with previous reports of damping-controlled spiral trajectories~\cite{Reichhardt2020_arXiv_MagnusClusters,Lin2013_PRB_ParticleModel,Buttner2015_NatPhys_SkyrmionDynamics}.

\begin{figure}[tb!]
    \centering
    \includegraphics[width=0.9\linewidth]{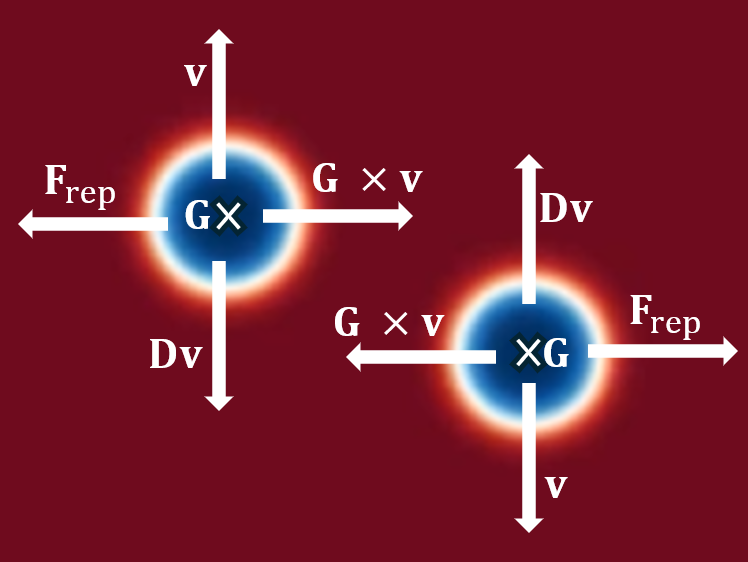}
\caption{
Schematic illustration of the forces, velocity and damping directions for two interacting skyrmions.}
    \label{fig:spiral_alpha}
\end{figure}

As the dynamics evolve, damping reduces the skyrmion velocity by weakening both the gyrotropic force and the skyrmion--skyrmion interaction force. This reduction leads to a gradual increase in the separation distance between the skyrmions~\cite{Reichhardt2020_arXiv_MagnusClusters,Lin2013_PRB_ParticleModel}. 
The force directions governing this behavior are depicted schematically in Fig.~\ref{fig:spiral_alpha}. For both skyrmions, the gyrovector points into the film along the $-z$ axis because the topological charge satisfies $Q>0$. The right skyrmion moves along $-y$ while the left skyrmion moves along $+y$. According to the right-hand rule, the Magnus force deflects the right skyrmion to the left and the left skyrmion to the right. Combined with the long-range repulsive force between the skyrmions, this deflection produces the characteristic clockwise spiral trajectory observed in the simulations~\cite{Capic2020_JPhysCondMat_SkyrmionInteraction,Brearton2020_arXiv_SkyrmionInteractions,Ross2021_SkyrmionInteractions,Reichhardt2020_arXiv_MagnusClusters,Lin2013_PRB_ParticleModel}. 

\begin{figure}[h!]
    \centering
    \includegraphics[width=1\linewidth]{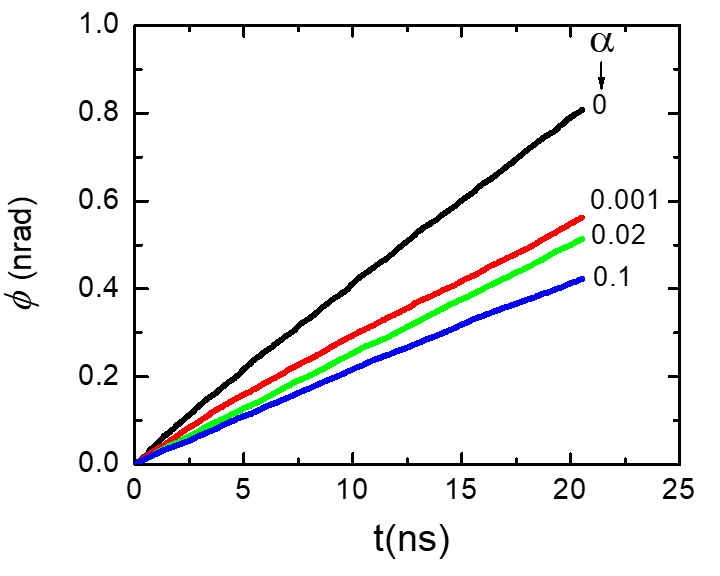}
\caption{
Time evolution of the angular position for two interacting skyrmions at different damping values $\alpha$.}
    \label{fig:FIG3}
\end{figure}

The damping dependence of the rotation is quantified in Fig.~\ref{fig:FIG3}, which shows the time evolution of the angular position for several damping values. The angular displacement increases approximately linearly with time, confirming a nearly constant angular velocity during the spiral motion. Lower damping ($\alpha = 0$ and $\alpha = 0.001$) results in higher angular velocities, whereas higher damping ($\alpha = 0.02$ and $\alpha = 0.1$) significantly suppresses the rotation rate, as expected from the Thiele formalism and previous numerical studies~\cite{Thiele1973_PRL_DomainMotion,Lin2013_PRB_ParticleModel,Reichhardt2020_arXiv_MagnusClusters}.

\begin{figure}[h!]
    \centering
    \includegraphics[width=1\linewidth]{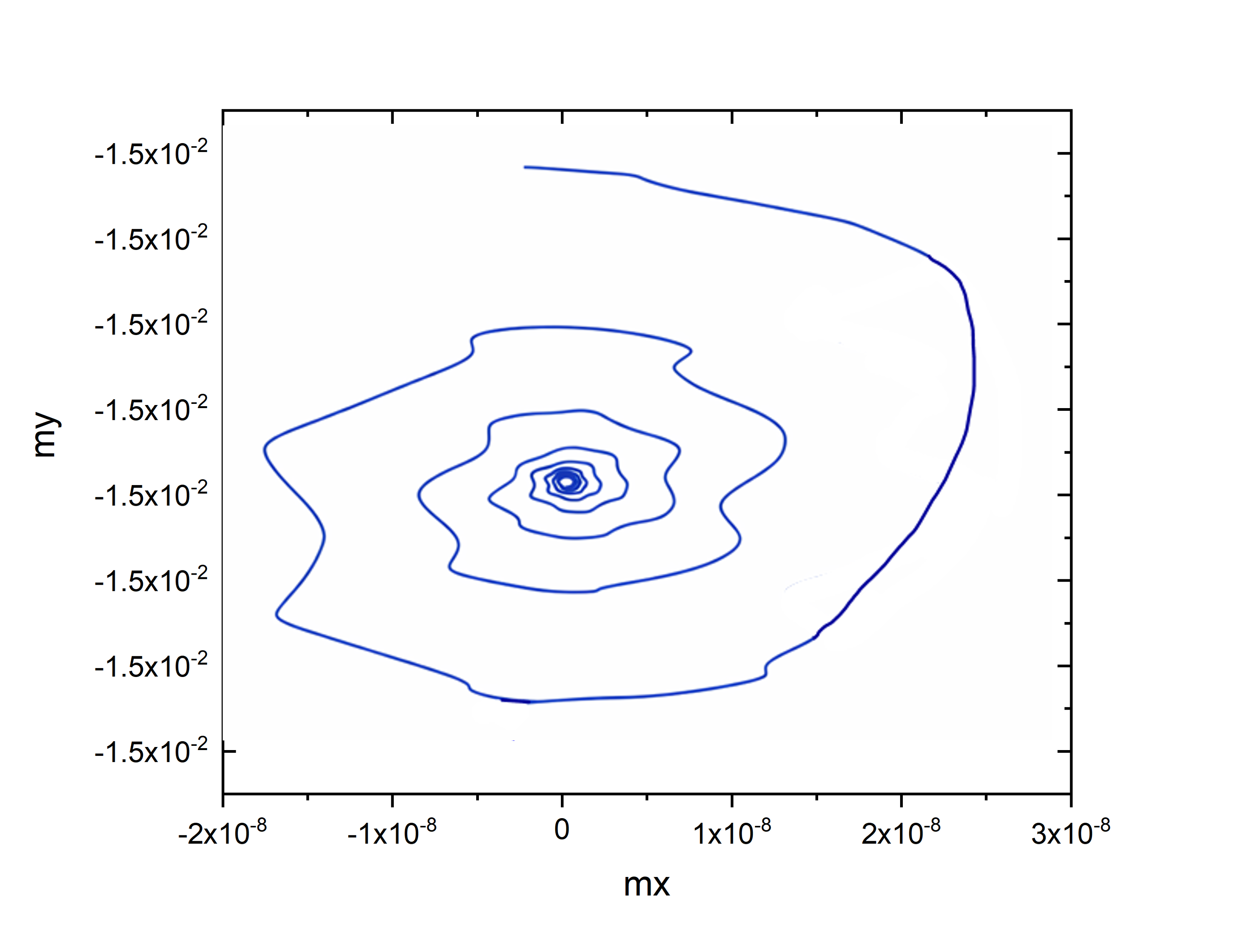}
\caption{Spiral trajectory of one skyrmion in the $x$--$y$ plane for $\alpha=0.1$.}
    \label{fig:spiral_plane}
\end{figure}

The trajectory shown in Fig.~4 corresponds to the in-plane magnetization components of a single isolated Néel skyrmion plotted in the $(m_x,m_y)$ plane. The motion forms a clockwise spiral in which $m_x$ oscillates noticeably while $m_y$ remains nearly constant. This behavior arises from the intrinsic gyrotropic precession of a skyrmion with topological charge $Q=1$, for which the gyrovector $G=-4\pi Q<0$ enforces clockwise rotation. Because a Néel skyrmion has a predominantly radial in-plane structure, the $m_x$ component is the one most sensitive to the periodic deformation associated with gyrotropic motion, whereas $m_y$ varies only weakly. The slow inward contraction of the spiral is caused by Gilbert damping, which gradually reduces the amplitude of the gyrotropic orbit in the absence of external forces or skyrmion–skyrmion interactions.
~\cite{Capic2020_JPhysCondMat_SkyrmionInteraction,Brearton2020_arXiv_SkyrmionInteractions,Reichhardt2020_arXiv_MagnusClusters,Lin2013_PRB_ParticleModel}.

\begin{figure}[h!]
    \centering
    \includegraphics[width=1\linewidth]{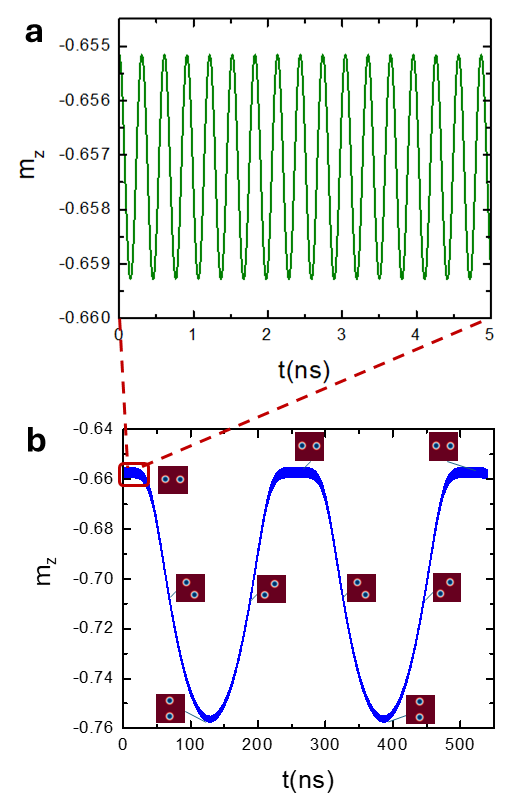}
\caption{Breathing and long-period oscillations of the skyrmion pair during spiral motion at $\alpha=0$.}
    \label{fig:FIG5}
\end{figure}

In Fig.~\ref{fig:FIG5}, for $\alpha = 0$, the out-of-plane magnetization $m_z(t)$ exhibits a two-component oscillatory behavior arising from the conservative skyrmion--skyrmion interaction. 
In the short-time dynamics shown in Fig.~\ref{fig:FIG5}(a) which is a zoom area in small portion of Fig.~\ref{fig:FIG5}(b), the signal exhibits a high-frequency breathing 
oscillation of small amplitude. This behaviour originates from the rapid periodic expansion and 
contraction of the skyrmion radius $R(t)$ due to the interaction force between the two skyrmions. 
As the radius varies, the spin configuration adjusts to accommodate the change in size, which 
modifies the out-of-plane component through the relation $m_{z}(r) = \cos\theta(r)$. This intrinsic breathing mode results from the internal exchange--DMI--anisotropy energy balance and appears at a frequency of about $3.2~\mathrm{GHz}$, consistent with previous studies of skyrmion breathing dynamics~\cite{Kim2014_PRB_BreathingSkyrmionDots,Garanin2020_BreathingMode,McKeever2019_PRB_BreathingDynamics,Mochizuki2012_PRL_SkyrmionSpinWave}. 

Over longer timescales, as shown in Fig.~\ref{fig:FIG5}(b), a low-frequency modulation of about $4~\mathrm{MHz}$ becomes visible, generating a large envelope superimposed on the fast oscillation. 
This slow component originates from the gradual spiral motion of the skyrmion pair: as their separation slowly changes along the spiral orbit, the average value of $m_z$ is modulated, producing a large-amplitude waveform. Because the damping vanishes ($\alpha = 0$), no energy is dissipated, and both the fast breathing oscillation and the slow spiral-motion modulation persist without decay. The pair completes a full $2\pi$ reorientation with a characteristic period of 
$T \approx 250\,\text{ns}$. Throughout this cycle, the out-of-plane magnetization 
of each skyrmion displays a pronounced oscillatory response: the core value 
$m_{z}$ varies between approximately $-0.66$ and $-0.76$ as the skyrmions rotate 
from a horizontal alignment to a vertical one. The resulting asymmetry between the upper and lower turning points reflects the anharmonic nature of the skyrmion--skyrmion potential and the nonuniform rate at which the spin texture reorganizes during the motion~\cite{McKeever2019_PRB_BreathingDynamics,Garanin2020_BreathingMode}.

In summary, $m_z(t)$ signal contains two well-separated frequencies: a fast breathing mode set by the intrinsic eigenfrequency of an individual skyrmion, and a slow modulation associated with the long-period spiral motion of the skyrmion pair~\cite{Kim2014_PRB_BreathingSkyrmionDots,Garanin2020_BreathingMode,McKeever2019_PRB_BreathingDynamics,Mochizuki2012_PRL_SkyrmionSpinWave}.

\begin{figure}[b!]
    \centering
    \includegraphics[width=1\linewidth]{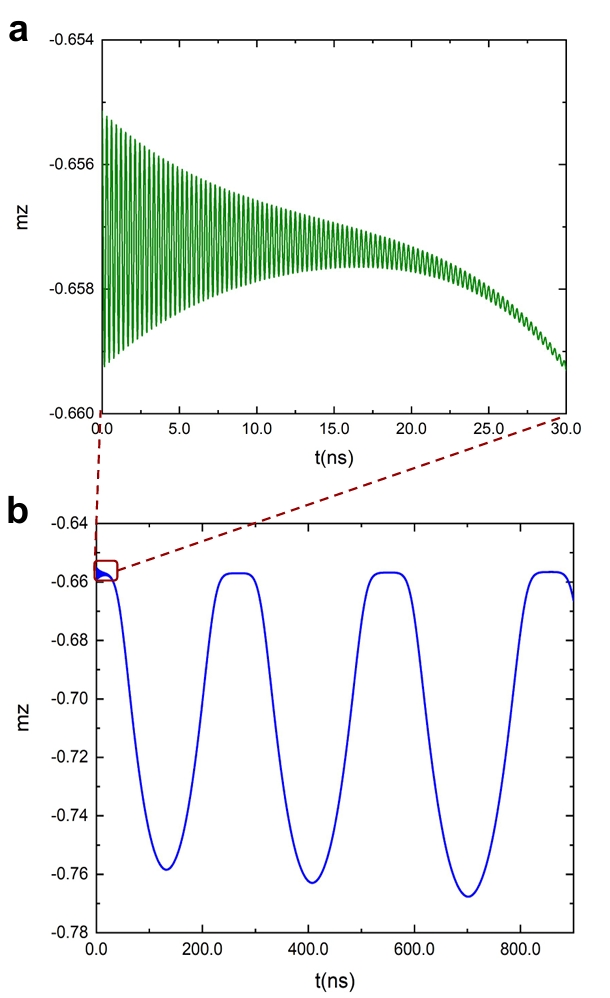}
\caption{Breathing and long-period oscillations of the skyrmion pair during spiral motion at $\alpha=0.001$.}
    \label{fig:FIG6}
\end{figure}

In Fig.~\ref{fig:FIG6}, when a small but finite damping is present ($\alpha=0.001$), the high-frequency breathing mode gradually decreases in amplitude and becomes fully damped after approximately $30~\mathrm{ns}$ [Fig.~\ref{fig:FIG6}(a)]. However, the long-period oscillation associated with the skyrmion spiral motion remains clearly visible and persists over the full simulation time [Fig.~\ref{fig:FIG6}(b)]. This behaviour is consistent with the picture that Gilbert damping primarily affects the internal eigenmodes, while the slow spiral motion is governed by the balance of gyrotropic and interaction forces captured by the Thiele equation~\cite{Thiele1973_PRL_DomainMotion,Lin2013_PRB_ParticleModel,Buttner2015_NatPhys_SkyrmionDynamics}.

\begin{figure}[b!]
    \centering
    \includegraphics[width=1\linewidth]{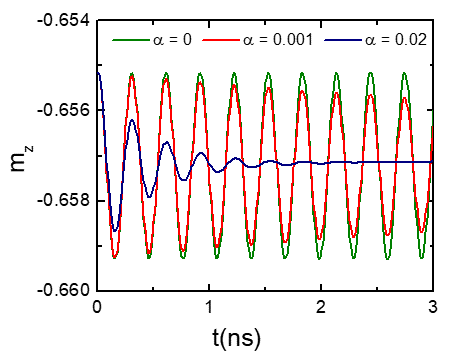}
\caption{The out-of-plane magnetization $m_z$ versus time for $\alpha=0$, $\alpha=0.001$, and $\alpha=0.02$.}
    \label{fig:FIG7}
\end{figure}

In Fig.~\ref{fig:FIG7}, the short-time evolution of the out-of-plane magnetization $m_z(t)$ is shown for two interacting skyrmions under different damping conditions, $\alpha = 0$, $\alpha = 0.001$, and $\alpha = 0.02$. In all cases, the system exhibits a high-frequency breathing oscillation with $f \approx 3.2~\mathrm{GHz}$. The frequency is essentially the same because it is set by the intrinsic skyrmion eigenmode rather than by damping. For $\alpha = 0$, the oscillation maintains a slightly larger amplitude with an almost perfectly periodic waveform, as no energy is lost during the motion. When a very small damping is introduced ($\alpha = 0.001$), the breathing frequency remains nearly unchanged, but the amplitude decreases slightly. Overall, the curves demonstrate that the breathing mode is an intrinsic, high-frequency component of skyrmion dynamics, while damping primarily influences the amplitude rather than the frequency~\cite{Garanin2020_BreathingMode,McKeever2019_PRB_BreathingDynamics,Kim2014_PRB_BreathingSkyrmionDots,Mochizuki2012_PRL_SkyrmionSpinWave}.

\subsection{Theoretical Arguments Supporting Numerical Results}

We start from Eq.~(6) and rewrite the velocity in polar coordinates following the standard collective-coordinate treatment of skyrmion motion~\cite{Thiele1973_PRL_DomainMotion,Lin2013_PRB_ParticleModel}. The unit vectors $\hat{\mathbf{r}}$, $\hat{\boldsymbol{\phi}}$ and $\hat{\mathbf{z}}$ are linked through
\begin{align}
\hat{\mathbf{z}} \times \hat{\mathbf{r}} &= \hat{\boldsymbol{\phi}} \\
\hat{\mathbf{z}} \times \hat{\boldsymbol{\phi}} &= -\hat{\mathbf{r}}
\end{align}

Substituting into the Thiele equation and separating components gives the radial and azimuthal components as expressed in the two following equations (12) and (13), respectively.

\begin{equation}
-G \dot{\phi} + \alpha D \dot{r} = F(r)
\end{equation}
\begin{equation}
G \dot{r} + \alpha D r \dot{\phi} = 0
\end{equation}
where $F(r) = -\frac{dU}{dr}$ is the radial interaction force. From these two equations, the radial velocity can be expressed as

\begin{equation}
\dot{r} = \frac{\alpha D}{G^{2} + (\alpha D)^{2}} \, F(r)
\end{equation}
Since the denominator is constant, $\dot{r}$ is thus proportional to 
 $\alpha D\, F(r)$. For $F(r) < 0$ case, the radial velocity becomes negative and the spiral motion is then inward. In contrast, if $F(r) > 0$, the spiral motion is outward.
 
The radial motion requires a material with a non-zero damping constant.  
Therefore, increasing $\alpha$ increases the radial velocity, as revealed by simulation results and reported by analytical predictions~\cite{Lin2013_PRB_ParticleModel,Reichhardt2020_arXiv_MagnusClusters}.

\subsection*{Integrating the Radial Motion}

Equation 12 can be written as 

\begin{equation}
\dot{r} = C\,F(r)
\end{equation}
where C is equal to $\alpha D / \bigl(G^{2} + (\alpha D)^{2}\bigr)$. This gives the separable differential equation
\begin{equation}
\frac{dr}{F(r)} = C\,dt
\end{equation}
With the initial condition $r(0)=r_{0}$, the motion of two skyrmions away from each other can be described by the autonomous equation
\begin{equation}
\int_{r_{0}}^{r(t)} \frac{dr'}{F(r')} 
= C t
\end{equation}

In our system, the interaction force decays exponentially at large separations, which leads to a 
logarithmic increase of the skyrmion--skyrmion distance over time, consistent with analytical forms of the interaction potential reported in Refs.~\cite{Capic2020_JPhysCondMat_SkyrmionInteraction,Ross2021_SkyrmionInteractions,Lin2013_PRB_ParticleModel}.

\subsection*{Angular Velocity}

Solving the azimuthal component gives
\begin{equation}
\dot{\phi} = -\frac{G}{r\left[G^{2} + (\alpha D)^{2}\right]} F(r)
\end{equation}
The sign of $\dot{\phi}$ is determined by $G \propto -Q$, which leads to clockwise (counterclockwise) rotation for $Q > 0$ ($Q < 0$). These results are in agreement with the handedness of skyrmion trajectories in chiral magnets~\cite{Nagaosa2013_NatNanotech,Lin2013_PRB_ParticleModel}. Its magnitude depends on $F(r)/r$ and increases as $\alpha$ is reduced (Fig.~\ref{fig:FIG3}).

\subsection*{Spiral Trajectory}

Dividing the two sides of Eq. (16) by $\dot{r}$ and integrating them from $\phi_{0}$ to $\phi$ gives

\begin{equation}
\phi(r) = \phi_{0} - \frac{G}{\alpha D}\ln\left(\frac{r}{r_{0}}\right)
\end{equation}

Equivalently,
\begin{equation}
r(\phi) = r_{0}\,
\exp\!\left[-\frac{\alpha D}{G}\left(\phi-\phi_{0}\right)\right]
\end{equation}
showing that each rotation decreases (or increases) the radius exponentially. For a full $2\pi$ rotation one obtains
\begin{equation}
r(2\pi)
= r_{0}\exp\!\left( -2\pi\frac{\alpha D}{G} \right)
\end{equation}
which provides the spiral contraction (or expansion) rate per cycle, in qualitative agreement with the trajectories observed in Fig.~\ref{fig:spiral_plane} and the particle-based analysis of Ref.~\cite{Lin2013_PRB_ParticleModel}.

\begin{figure}[b!]
    \centering
    \includegraphics[width=0.7\linewidth]{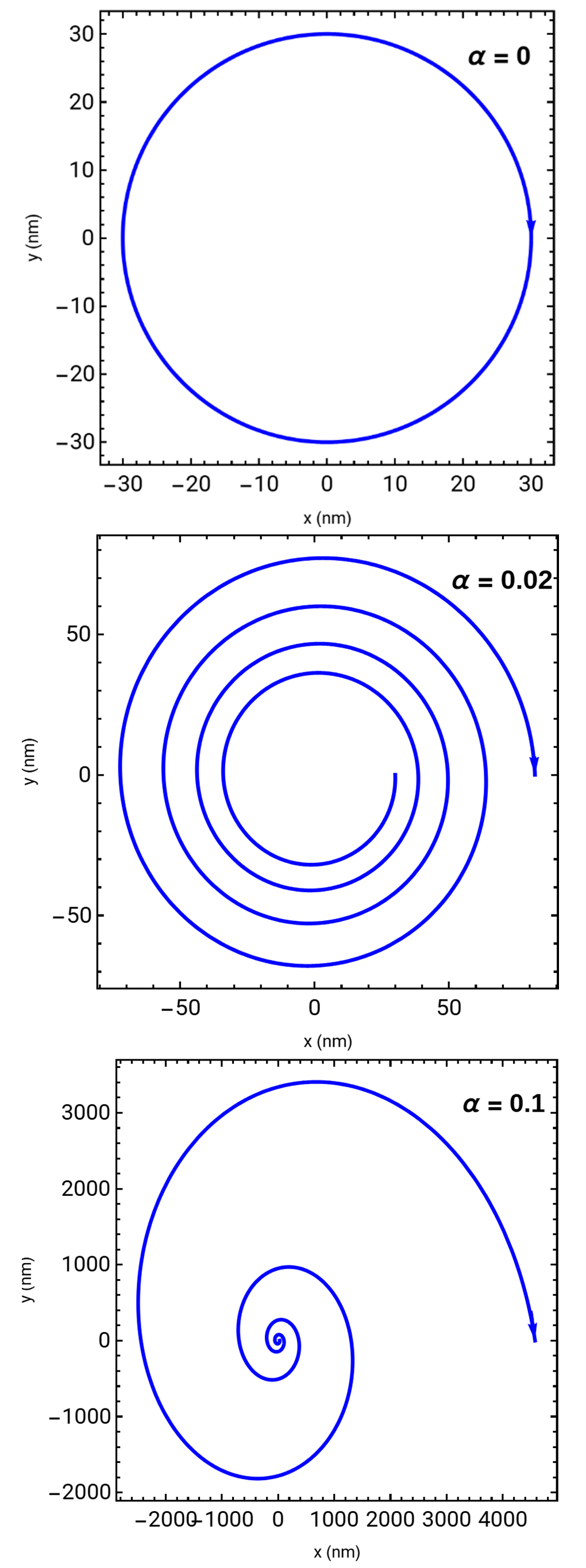}
\caption{Spiral trajectory of the skyrmion obtained from the analytical Eq.~(26), illustrating the gradual radial contraction with increasing rotation angle.}
    \label{fig:spiral_analytic}
\end{figure}

\subsection*{Long-Distance Behavior of Skyrmions}

To analyze the asymptotic behavior of isolated skyrmions, we used the reduced micromagnetic functional in which dipolar interactions, external fields, and the Dzyaloshinskii--Moriya interaction become negligible at large distances~\cite{Capic2020_JPhysCondMat_SkyrmionInteraction,Ross2021_SkyrmionInteractions}. In this limit, the energy is reduced to
\begin{equation}
E = \int d^{2}r\,
\left[
A (\nabla \mathbf{s})^{2}
+ K_{\mathrm{eff}} \left(1 - s_{z}^{2}\right)
\right]
\label{eq:reduced_energy}
\end{equation}
where $\mathbf{s}=(s_{x},s_{y},s_{z})$ is the unit magnetization vector.

Far from the skyrmion core, the magnetisation is nearly uniform; i.e, $s_{x}, s_{y} \ll 1$ and $s_{z} \approx 1$,
which gives the small-angle approximation
Substituting into Eq.~\eqref{eq:reduced_energy} and keeping terms up to second order leads to
\begin{equation}
E = \int d^{2}r
\left[
A\!\left((\nabla s_{x})^{2}+(\nabla s_{y})^{2}\right)
+ \frac{K_{\mathrm{eff}}}{2}(s_{x}^{2}+s_{y}^{2})
\right]
\label{eq:quadratic_energy}
\end{equation}
which coincides with the quadratic exchange--anisotropy functional used in Refs.~\cite{Capic2020_JPhysCondMat_SkyrmionInteraction,Ross2021_SkyrmionInteractions}.
Variation of Eq.~\eqref{eq:quadratic_energy} yields the Helmholtz-type equation
\begin{equation}
-2A\nabla^{2}s_{i} + K_{\mathrm{eff}} s_{i} = 0,
\qquad i=x,y
\label{eq:Helmholtz}
\end{equation}
describing small transverse fluctuations on top of a uniform ferromagnetic background~\cite{Nagaosa2013_NatNanotech,Ross2021_SkyrmionInteractions}. For a circular N\'eel-type structure we use
\begin{equation}
s_{x}=S_{\perp}(r)\cos\varphi,
\qquad
s_{y}=S_{\perp}(r)\sin\varphi
\end{equation}
which gives the radial equation
\begin{equation}
S_{\perp}''(r)
+ \frac{1}{r} S_{\perp}'(r)
- \left(\frac{1}{r^{2}} + \frac{K_{\mathrm{eff}}}{2A}\right)
S_{\perp}(r)=0
\label{eq:radial}
\end{equation}

\begin{figure}[t!]
    \centering
    \includegraphics[width=1.0\linewidth]{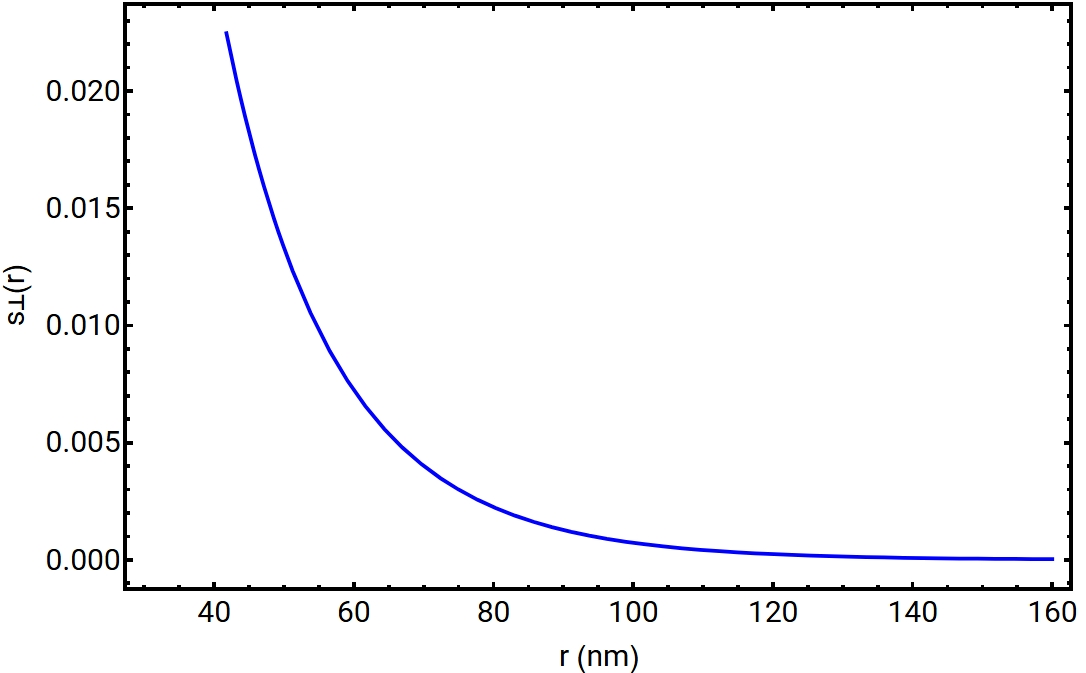}
\caption{Exponential decay of the out-of-plane spin component $s_{\perp}(r)$,
consistent with the asymptotic form obtained from Eq.~(24).}
    \label{fig:kdecay}
\end{figure}

Defining the decay length $\delta_{k}$ as $\sqrt{A/K_{\mathrm{eff}}}$,
Eq.~\eqref{eq:radial} becomes the modified Bessel equation with solution
\begin{equation}
S_{\perp}(r)=C_{1} I_{1}\!\left(\frac{r}{\delta_{k}}\right)
+ C_{2} K_{1}\!\left(\frac{r}{\delta_{k}}\right)
\end{equation}
The physical decay requires $C_{1}=0$, giving the long-distance tail
\begin{equation}
s_{\perp}(r)=C\, K_{1}\!\left(\frac{r}{\delta_{k}}\right)
\label{eq:tail_solution}
\end{equation}
with asymptotic limits
\begin{align}
r \ll \delta_{k}
&:\quad s_{\perp}(r)\sim \frac{1}{r} \\[3pt]
r \gg \delta_{k}
&:\quad s_{\perp}(r)\sim
C\sqrt{\frac{\pi\delta_{k}}{2r}}
\exp\!\left(-\frac{r}{\delta_{k}}\right)
\end{align}
The overlap between two skyrmion tails therefore decreases as $\exp(-r/\delta_{k})$, implying an exponentially decaying long-range skyrmion--skyrmion interaction, as shown in Fig.~\ref{fig:kdecay} and reported in Refs.~\cite{Capic2020_JPhysCondMat_SkyrmionInteraction,Ross2021_SkyrmionInteractions}.

\section{Conclusion}

In this work, we presented a numerical and analytical study of the interaction dynamics of two Néel skyrmions in an ultrathin ferromagnetic film. Micromagnetic simulations revealed that skyrmion pairs with identical polarity and opposite chirality exhibit a long-range repulsive interaction when their initial separation exceeds twice the skyrmion radius. This repulsion drives an outward spiral motion whose angular and radial components depend sensitively on the Gilbert damping constant. In the undamped limit, the dynamics are dominated by the gyrotropic force, producing nearly circular orbits accompanied by a persistent high-frequency breathing mode. When damping is introduced, the angular velocity decreases, the radial drift increases, and the breathing oscillations are gradually suppressed, leaving a slow modulation associated with the evolving skyrmion separation.  

To interpret these results, we used the Thiele collective-coordinate model, which accurately captures the spiral trajectories, the logarithmic increase of the separation distance, and the damping dependence of both radial and angular velocities. The analytical expressions derived from the Thiele equation show excellent agreement with the simulated trajectories. Furthermore, a continuum micromagnetic analysis of the skyrmion tail demonstrates that the in-plane magnetization decays rapidly at large distances, explaining the exponential form of the long-range interaction potential observed in both the numerical and analytical models.  

Overall, this multiscale approach establishes a unified physical picture linking skyrmion topology, long-range repulsion, gyrotropic motion, and internal breathing dynamics. The insights of the study provide a fundamental understanding of skyrmion–skyrmion interactions and offer guidance for designing future skyrmion-based information carriers, devices relying on collective skyrmion motion, and architectures where controlled multi-skyrmion dynamics are essential.


\bibliographystyle{apsrev4-2}
\bibliography{references}

@article{Fert2017_NatRevMat_Skyrmions,
  author  = {Fert, A. and Reyren, N. and Cros, V.},
  title   = {Magnetic skyrmions: advances in physics and applications},
  journal = {Nature Reviews Materials},
  volume  = {2},
  pages   = {17031},
  year    = {2017},
  doi     = {10.1038/natrevmats.2017.31},
  url     = {https://doi.org/10.1038/natrevmats.2017.31}
}

@article{Nagaosa2013_NatNanotech,
  author  = {Nagaosa, Naoto and Tokura, Yoshinori},
  title   = {Topological properties and dynamics of magnetic skyrmions},
  journal = {Nat. Nanotechnol.},
  volume  = {8},
  pages   = {899--911},
  year    = {2013},
  doi     = {10.1038/nnano.2013.243}
}

@article{Woo2016_NatMat_RoomTempSkyrmions,
  author  = {Woo, S. and Litzius, K. and Kr{\"u}ger, B. and et al.},
  title   = {Observation of room-temperature magnetic skyrmions and their current-driven dynamics},
  journal = {Nature Materials},
  volume  = {15},
  pages   = {501--506},
  year    = {2016},
  doi     = {10.1038/nmat4593},
  url     = {https://doi.org/10.1038/nmat4593}
}

@article{Mochizuki2012_PRL_SkyrmionSpinWave,
  author  = {Mochizuki, M.},
  title   = {Spin-wave modes and their intense excitation effects in skyrmion crystals},
  journal = {Physical Review Letters},
  volume  = {108},
  pages   = {017601},
  year    = {2012},
  doi     = {10.1103/PhysRevLett.108.017601},
  url     = {https://doi.org/10.1103/PhysRevLett.108.017601}
}

@article{Thiele1973_PRL_DomainMotion,
  author  = {Thiele, A. A.},
  title   = {Steady-state motion of magnetic domains},
  journal = {Physical Review Letters},
  volume  = {30},
  pages   = {230--233},
  year    = {1973},
  doi     = {10.1103/PhysRevLett.30.230},
  url     = {https://doi.org/10.1103/PhysRevLett.30.230}
}

@article{Lin2013_PRB_ParticleModel,
  author  = {Lin, S.-Z. and Reichhardt, C. and Batista, C. D. and Saxena, A.},
  title   = {Particle model for skyrmions in metallic chiral magnets: Dynamics, pinning, and creep},
  journal = {Physical Review B},
  volume  = {87},
  pages   = {214419},
  year    = {2013},
  doi     = {10.1103/PhysRevB.87.214419},
  url     = {https://doi.org/10.1103/PhysRevB.87.214419}
}

@article{Shi2023_JPhysD_SkyrmionCollision,
  author  = {Shi, S. and Zhao, Y. and Sun, J. and Hou, X. and Zhou, H. and Wang, J.},
  title   = {Dynamic behavior of skyrmion collision: spiral and breath},
  journal = {Journal of Physics D: Applied Physics},
  volume  = {56},
  pages   = {485001},
  year    = {2023},
  doi     = {10.1088/1361-6463/aceb3a},
  url     = {https://doi.org/10.1088/1361-6463/aceb3a}
}

@article{Reichhardt2020_arXiv_MagnusClusters,
  author  = {Reichhardt, C. and Reichhardt, C. J. O.},
  title   = {Dynamics of Magnus dominated particle clusters, collisions, pinning and ratchets},
  journal = {arXiv},
  year    = {2020},
  eprint  = {2002.09794},
  url     = {https://arxiv.org/abs/2002.09794}
}

@article{Buttner2015_NatPhys_SkyrmionDynamics,
  author  = {B{\"u}ttner, F. and Moutafis, C. and Schmidt, M. and et al.},
  title   = {Dynamics and inertia of skyrmionic spin structures},
  journal = {Nature Physics},
  volume  = {11},
  pages   = {225--228},
  year    = {2015},
  doi     = {10.1038/nphys3231},
  url     = {https://doi.org/10.1038/nphys3231}
}

@article{Vansteenkiste2014_AIPAdv_MuMax3,
  author  = {Vansteenkiste, A. and Leliaert, J. and Dvornik, M. and Helsen, M. and Garcia-Sanchez, F. and Van Waeyenberge, B.},
  title   = {The design and verification of MuMax3},
  journal = {AIP Advances},
  volume  = {4},
  pages   = {107133},
  year    = {2014},
  doi     = {10.1063/1.4899186},
  url     = {https://doi.org/10.1063/1.4899186}
}

@article{Brearton2020_arXiv_SkyrmionInteractions,
  author  = {Brearton, R. and van der Laan, G. and Hesjedal, T.},
  title   = {Magnetic skyrmion interactions in the micromagnetic framework},
  journal = {arXiv},
  year    = {2020},
  eprint  = {2001.08218},
  url     = {https://arxiv.org/abs/2001.08218}
}

@article{AlSubhi2019_CoNi,
  author  = {Al Subhi, A. and Sbiaa, R.},
  title   = {Control of magnetization reversal and domain structure in (Co/Ni) multilayers},
  journal = {Journal of Magnetism and Magnetic Materials},
  volume  = {489},
  pages   = {165460},
  year    = {2019},
  doi     = {10.1016/j.jmmm.2019.165460},
  url     = {https://doi.org/10.1016/j.jmmm.2019.165460}
}

@article{Sbiaa2016_CoNiCoPt_FMR,
  author  = {Sbiaa, R. and Shaw, J. M. and Nembach, H. T. and Al Bahri, M. and Ranjbar, M. and Åkerman, J. and Piramanayagam, S. N.},
  title   = {Ferromagnetic resonance measurements of (Co/Ni/Co/Pt) multilayers with perpendicular magnetic anisotropy},
  journal = {Journal of Physics D: Applied Physics},
  volume  = {49},
  number  = {42},
  pages   = {425002},
  year    = {2016},
  doi     = {10.1088/0022-3727/49/42/425002},
  url     = {https://doi.org/10.1088/0022-3727/49/42/425002}
}

@article{Tomasello2014_SciRep_SkyrmionRacetrack,
  author  = {Tomasello, R. and Martinez, E. and Zivieri, R. and Torres, L. and Carpentieri, M. and Finocchio, G.},
  title   = {A strategy for the design of skyrmion racetrack memories},
  journal = {Scientific Reports},
  volume  = {4},
  pages   = {6784},
  year    = {2014},
  doi     = {10.1038/srep06784},
  url     = {https://doi.org/10.1038/srep06784}
}

@article{Song2017_SkyrmionRacetrack,
  author  = {Song, C. and Jin, C. and Wang, J. and Xia, H. and Wang, J. and Liu, Q.},
  title   = {Skyrmion-based multi-channel racetrack},
  journal = {Applied Physics Letters},
  volume  = {111},
  number  = {19},
  pages   = {192413},
  year    = {2017},
  doi     = {10.1063/1.5001457},
  url     = {https://doi.org/10.1063/1.5001457}
}

@article{Bessarab2018_SkyrmionLifetime,
  author  = {Bessarab, P. F. and M{\"u}ller, G. P. and Lobanov, I. S. and Mryasov, O. N. and Kiselev, N. S. and Stamps, R. L.},
  title   = {Lifetime of racetrack skyrmions},
  journal = {Physical Review B},
  volume  = {99},
  number  = {5},
  pages   = {054420},
  year    = {2019},
  doi     = {10.1103/PhysRevB.99.054420},
  url     = {https://doi.org/10.1103/PhysRevB.99.054420}
}

@article{Huang2020_NatElectron_SkyrmionSynapse,
  author  = {Huang, Y. and Kang, W. and Zhang, X. and Zhou, Y. and Zhao, W.},
  title   = {Artificial synapse based on skyrmion motion controlled by magnetic field gradient},
  journal = {Nature Electronics},
  volume  = {3},
  pages   = {122--130},
  year    = {2020},
  doi     = {10.1038/s41928-020-0371-4},
  url     = {https://doi.org/10.1038/s41928-020-0371-4}
}

@article{He2023_NanoLett_SkyrmionMemory,
  author  = {He, B. and Tomasello, R. and Luo, X. and Zhang, R. and Nie, Z. and Carpentieri, M. and Han, X. and Finocchio, G. and Yu, G.},
  title   = {All-electrical 9-bit skyrmion-based racetrack memory designed with laser irradiation},
  journal = {Nano Letters},
  volume  = {23},
  pages   = {9482--9490},
  year    = {2023},
  doi     = {10.1021/acs.nanolett.3c02036},
  url     = {https://doi.org/10.1021/acs.nanolett.3c02036}
}

@article{Pham2024_Science_SkyrmionSAF,
  author  = {Pham, V. T. and Sisodia, H. and Takahashi, Y. K. and Sugimoto, S. and Mitani, S.},
  title   = {Fast current-induced skyrmion motion in synthetic antiferromagnets},
  journal = {Science},
  volume  = {384},
  pages   = {307--312},
  year    = {2024},
  doi     = {10.1126/science.add5751},
  url     = {https://doi.org/10.1126/science.add5751}
}

@article{Zhang2015_SciRep_SkyrmionLogic,
  author  = {Zhang, X. and Ezawa, M. and Zhou, Y.},
  title   = {Magnetic skyrmion logic gates: conversion, duplication and merging of skyrmions},
  journal = {Scientific Reports},
  volume  = {5},
  pages   = {9400},
  year    = {2015},
  doi     = {10.1038/srep09400},
  url     = {https://doi.org/10.1038/srep09400}
}

@article{Mankalale2019_IEEE_SkyLogic,
  author  = {Mankalale, M. G. and Saha, S. and Nam, P. N. T. and Li, H. and Toroczkai, Z.},
  title   = {SkyLogic—A proposal for a skyrmion-based logic device},
  journal = {IEEE Transactions on Electron Devices},
  volume  = {66},
  pages   = {1990--1996},
  year    = {2019},
  doi     = {10.1109/TED.2019.2903267},
  url     = {https://doi.org/10.1109/TED.2019.2903267}
}

@article{Yu2020NanoscaleAdv,
  author  = {Yu, Ziyang and Shen, Maokang and Zeng, Zhongming and Liang, Shiheng
             and Liu, Yong and Chen, Ming and Zhang, Zhenhua and Lu, Zhihong
             and You, Long and Yang, Xiaofei and Zhang, Yue and Xiong, Rui},
  title   = {Voltage-controlled skyrmion-based nanodevices for neuromorphic computing using a synthetic antiferromagnet},
  journal = {Nanoscale Adv.},
  volume  = {2},
  pages   = {1309--1317},
  year    = {2020},
  doi     = {10.1039/D0NA00009D}
}

@article{Sisodia2022_PRApp_SkyrmionLogic,
  author  = {Sisodia, N. and Wang, J. and Zhang, X. and Zhou, Y.},
  title   = {Programmable skyrmion logic gates based on skyrmion tunneling},
  journal = {Physical Review Applied},
  volume  = {17},
  pages   = {064035},
  year    = {2022},
  doi     = {10.1103/PhysRevApplied.17.064035},
  url     = {https://doi.org/10.1103/PhysRevApplied.17.064035}
}

@article{MousaviCheghabouri2024_AdvTheorySimul_SkyrmionLogic,
  author  = {Mousavi Cheghabouri, A. and Yagan, R. and Onbasi, M.},
  title   = {Scalable low-power skyrmionic logic gate library},
  journal = {Advanced Theory and Simulations},
  volume  = {7},
  pages   = {2400243},
  year    = {2024},
  doi     = {10.1002/adts.202400243},
  url     = {https://doi.org/10.1002/adts.202400243}
}

@article{Belrhazi2024_ACSApplMater_SkyrmionLogic,
  author  = {Belrhazi, H. and Fattouhi, M. and El Hafidi, M. Y. and El Hafidi, M.},
  title   = {Reconfigurable skyrmion-based logic gates: versatile design and full-scale implementation},
  journal = {ACS Applied Materials \& Interfaces},
  volume  = {16},
  pages   = {3703--3718},
  year    = {2024},
  doi     = {10.1021/acsami.3c16542},
  url     = {https://doi.org/10.1021/acsami.3c16542}
}

@article{Pan2024_PRResearch_SkyrmionQubits,
  author  = {Pan, X. F. and Hei, X. L. and Yao, X. Y. and Chen, J. Q. and Ren, Y. M. and Dong, X. L. and Li, P. B. and et al.},
  title   = {Skyrmion-mechanical hybrid quantum systems: manipulation of skyrmion qubits via phonons},
  journal = {Physical Review Research},
  volume  = {6},
  pages   = {023067},
  year    = {2024},
  doi     = {10.1103/PhysRevResearch.6.023067},
  url     = {https://doi.org/10.1103/PhysRevResearch.6.023067}
}

@article{Lone2024_Nanoscale_MultilayerSpintronic,
  author  = {Lone, A. H. and Zou, Xuecui and Mishra, Kishan K. and Singaravelu, Venkatesh and Sbiaa, R. and Fariborzi, Hossein and Setti, Gianluca},
  title   = {Multilayer ferromagnetic spintronic devices for neuromorphic computing applications},
  journal = {Nanoscale},
  volume  = {16},
  pages   = {12431--12444},
  year    = {2024},
  doi     = {10.1039/D4NR01003E},
  url     = {https://doi.org/10.1039/D4NR01003E}
}

@article{AlSaidi2025_MatTodayE_SkyrmionBiskyrmion,
  author  = {Al Saidi, W. and Amara, S. and Zar Myint, M. T. and Al Harthi, S. and Setti, G. and Sbiaa, R.},
  title   = {Stabilized biskyrmion states in annealed CoFeB bilayer with different interfaces},
  journal = {Materials Today Electronics},
  volume  = {13},
  pages   = {100166},
  year    = {2025},
  doi     = {10.1016/j.mtelec.2025.100166},
  url     = {https://doi.org/10.1016/j.mtelec.2025.100166}
}

@article{Verma2025_APL_FieldFreeSkyrmionNN,
  author  = {Verma, Shubhi and Khosla, Aman and Medwal, Rohit and Ojha, Animesh K. and Gupta, Surbhi},
  title   = {Biased field-free skyrmion-based neural network and reconfigurable logic operations},
  journal = {Applied Physics Letters},
  volume  = {127},
  number  = {20},
  pages   = {202410},
  year    = {2025},
  doi     = {10.1063/5.0291591},
  url     = {https://doi.org/10.1063/5.0291591}
}

@article{Song2020_NatElectron_SkyrmionSynapses,
  author  = {Song, Kyung Mee and Jeong, Jae-Seung and Pan, Biao and Zhang, Xichao and Xia, Jing and Cha, Sunkyung and Park, Tae-Eon and Kim, Kwangsu and Finizio, Simone and Raabe, Jörg and Chang, Joonyeon and Zhou, Yan and Zhao, Weisheng and Kang, Wang and Ju, Hyunsu and Woo, Seonghoon},
  title   = {Skyrmion-based artificial synapses for neuromorphic computing},
  journal = {Nature Electronics},
  volume  = {3},
  pages   = {148--155},
  year    = {2020},
  doi     = {10.1038/s41928-020-0385-0},
  url     = {https://doi.org/10.1038/s41928-020-0385-0}
}

@article{Zhang2018_NanoLett_SkyrmionNeuron,
  author  = {Zhang, X. and Zhu, W. and Wang, Y. and Zhong, H.
             and Xia, J. and Zhou, Y. and Zhao, W.},
  title   = {Skyrmion-based artificial neuron device},
  journal = {Nano Lett.},
  volume  = {18},
  pages   = {1057--1063},
  year    = {2018},
  doi     = {10.1021/acs.nanolett.7b04722}
}

@article{Aslam2025_ACSAELM_SkyrmionMemristor,
  author  = {Aslam, M. and Li, Q. and Wu, J. and Chen, T. and Gupta, S. and Lee, D.},
  title   = {Low‑power skyrmion memristors and reservoir computing for neuromorphic hardware},
  journal = {ACS Applied Electronic Materials},
  volume  = {7},
  pages   = {1234--1245},
  year    = {2025},
  doi     = {10.1021/acsaelm.4c01827},
  url     = {https://doi.org/10.1021/acsaelm.4c01827}
}

@article{Bo2023_PRB_SkyrmionBags,
  author  = {Bo, Lan and Zhao, Rongzhi and Hu, Chenglong and Zhang, Xichao and Zhang, Xuefeng and Mochizuki, Masahito},
  title   = {Controllable creation of skyrmion bags in a ferromagnetic nanodisk},
  journal = {Physical Review B},
  volume  = {107},
  pages   = {224431},
  year    = {2023},
  doi     = {10.1103/PhysRevB.107.224431},
  url     = {https://doi.org/10.1103/PhysRevB.107.224431}
}

@article{Wang2023_PRB_ParticleContinuumSkyrmions,
  author  = {Wang, X. R. and Hu, X. C.},
  title   = {Particle‑continuum duality of skyrmions},
  journal = {Physical Review B},
  volume  = {107},
  pages   = {174412},
  year    = {2023},
  doi     = {10.1103/PhysRevB.107.174412},
  url     = {https://doi.org/10.1103/PhysRevB.107.174412}
}

@article{Tang2021_NatNanotech_SkyrmionBundles,
  author  = {Tang, Jin and Wu, Yaodong and Wang, Weiwei and Kong, Lingyao and Lv, Boyao and Wei, Wensen and Zang, Jiadong and Tian, Mingliang and Du, Haifeng},
  title   = {Magnetic skyrmion bundles and their current‑driven dynamics},
  journal = {Nature Nanotechnology},
  volume  = {16},
  pages   = {1086--1091},
  year    = {2021},
  doi     = {10.1038/s41565-021-00954-9},
  url     = {https://doi.org/10.1038/s41565-021-00954-9}
}

@article{Yu2014_NatCommun_Biskyrmions,
  author  = {Yu, X. Z. and Tokunaga, Y. and Kaneko, Y. and Zhang, W. Z. and Kimoto, K. and Matsui, Y. and Taguchi, Y. and Tokura, Y.},
  title   = {Biskyrmion states and their current‑driven motion in a layered manganite},
  journal = {Nature Communications},
  volume  = {5},
  pages   = {3198},
  year    = {2014},
  doi     = {10.1038/ncomms4198},
  url     = {https://doi.org/10.1038/ncomms4198}
}

@article{Song2023_SciAdv_Biskyrmions,
  author  = {Song, Yuzhu and Xu, Tiankuo and Zhao, Guoping and Xu, Yuanji and Zhong, Zhicheng and Zheng, Xinqi and Shi, Naike and Zhou, Chang and Hao, Yiqing and Huang, Qingzhen and Xing, Xianran and Zhang, Ying and Chen, Jun},
  title   = {High‑density, spontaneous magnetic biskyrmions induced by negative thermal expansion in ferrimagnets},
  journal = {Science Advances},
  volume  = {9},
  pages   = {eadi1984},
  year    = {2023},
  doi     = {10.1126/sciadv.adi1984},
  url     = {https://doi.org/10.1126/sciadv.adi1984}
}

@article{Garanin2020_BreathingMode,
  author  = {Garanin, D.\ A. and Jaafar, R. and Chudnovsky, E.\ M.},
  title   = {Breathing mode of a skyrmion on a lattice},
  journal = {Phys. Rev. B},
  volume  = {101},
  pages   = {014418},
  year    = {2020},
  doi     = {10.1103/PhysRevB.101.014418},
  url     = {https://doi.org/10.1103/PhysRevB.101.014418}
}

@article{Gobel2019_Biskyrmions,
  author  = {G{\"o}bel, B. and Henk, J. and Mertig, I.},
  title   = {Forming individual magnetic biskyrmions by merging two skyrmions in a centrosymmetric nanodisk},
  journal = {Sci. Rep.},
  volume  = {9},
  pages   = {9521},
  year    = {2019},
  doi     = {10.1038/s41598-019-45965-8},
  url     = {https://doi.org/10.1038/s41598-019-45965-8}
}

@article{Ross2021_SkyrmionInteractions,
  author  = {Ross, C. and Sakai, N. and Nitta, M.},
  title   = {Skyrmion interactions and lattices in chiral magnets: analytical results},
  journal = {J. High Energy Phys.},
  volume  = {2021},
  number  = {2},
  pages   = {095},
  year    = {2021},
  doi     = {10.1007/JHEP02(2021)095},
  url     = {https://doi.org/10.1007/JHEP02(2021)095}
}

@article{Capic2020_JPhysCondMat_SkyrmionInteraction,
  author  = {Capic, D. A. and Garanin, D. A. and Chudnovsky, E. M.},
  title   = {Skyrmion–skyrmion interaction in a magnetic film},
  journal = {Journal of Physics: Condensed Matter},
  volume  = {32},
  pages   = {415803},
  year    = {2020},
  doi     = {10.1088/1361-648X/ab9bc8},
  url     = {https://doi.org/10.1088/1361-648X/ab9bc8}
}

@article{Kim2014_PRB_BreathingSkyrmionDots,
  author  = {Kim, J.-V. and Garcia-Sanchez, F. and Sampaio, J. and Moreau-Luchaire, C. and Cros, V. and Fert, A.},
  title   = {Breathing modes of confined skyrmions in ultrathin magnetic dots},
  journal = {Physical Review B},
  volume  = {90},
  pages   = {064410},
  year    = {2014},
  doi     = {10.1103/PhysRevB.90.064410},
  url     = {https://doi.org/10.1103/PhysRevB.90.064410}
}

@article{McKeever2019_PRB_BreathingDynamics,
  author  = {McKeever, B. F. and Rodrigues, D. R. and Pinna, D. and Abanov, A. and Sinova, J. and Everschor-Sitte, K.},
  title   = {Characterizing breathing dynamics of magnetic skyrmions and antiskyrmions within the Hamiltonian formalism},
  journal = {Physical Review B},
  volume  = {99},
  pages   = {054430},
  year    = {2019},
  doi     = {10.1103/PhysRevB.99.054430},
  url     = {https://doi.org/10.1103/PhysRevB.99.054430}
}

\end{document}